# Model of Multilayer Knowledge Diffusion for Competence Development in an Organization

Przemysław Różewski (*), Jarosław Jankowski

Faculty of Computer Science and Information Technology,

West Pomeranian University of Technology,

ul. Żołnierska 49, 71-210 Szczecin, Poland

***Abstract***: *Growing role of intellectual capital within organizations is affecting new strategies related to knowledge management and competence development. Among different aspects related to this field, knowledge diffusion has become one of interesting areas from both practitioner and researcher's perspective. Several models were proposed with main goal to simulate diffusion and to explain the nature of these processes. Existing models are focused on knowledge diffusion and they assume diffusion within a single layer using knowledge representation. From the organizational perspective connecting several types of knowledge and modelling changes of competence can bring additional value. In the article we extended existing approaches by using multilayer diffusion model and focused on analysis of competence development process. The proposed model describes competence development process in a new way through horizontal and vertical knowledge diffusion in multilayer network. In the network, agents collaborate and interchange various kind of knowledge through different layers and this mutual activities affect the competences in a positive or negative way. Taking under consideration worker's cognitive and social abilities and the previous level of competence the* *new competence level can be estimated. The model is developed to support competence management in different organizations.*

(*) Corresponding author (P. Różewski): prozewski@wi.zut.edu.pl,

# 1   Introduction

Employees' competence become the main part of organization's intellectual capital [1]. According to [2] the management and control of knowledge and skills, and more recently the management of organizations' competencies have turned out to be essential factors of industrial processes' performance. Modern companies are no longer production systems of products and services but create and sell knowledge-based products. Including competence management into production process required integrating new decision processes regarding the cognitive dimension of business, at every managerial level [2]. Moreover, the companies have to expand knowledge management to competence management. As a result the companies will be able to fulfil the following items [3]: find the right single employee for a specific task or project, retrieve and assemble flexible project teams, develop and update employees skills, explore the employees future career paths, speed up innovation management. The workers become knowledge workers [4] and continuing needs for upgrading workplace knowledge, skills and competencies is developed. Changes in work and the ways in which it is carried out bring the need for upgrading workplace knowledge, skills and competencies. In today's workplaces, and for a number of reasons, workloads are higher than ever and stress is a growing concern [5].

Competence is an observable or measurable ability of an actor to perform a necessary action(s) in a given context(s) to achieve a specific outcome(s) [6]. After analysis of various competence definitions [7,8] one thing is common, competence is made of different knowledge-based components (e.g. knowledge, skills, behaviours). Competence development process is an acquisition of a specific set of competence's components that constitutes a particular competence [9].

In our work the modelling of the competence development process is based on the knowledge diffusion model that extends current solutions. The approach is new and required special characteristics of diffusion model. We developed a multilayer diffusion model based on the multilayer graph reflecting organisation's network. In the graph each layer represents competence's component (some kind of knowledge). There is an interaction between layers defined as a vertical diffusion. The horizontal diffusion occurs on every layer's level and relates to the diffusion of one type of knowledge between knowledge workers. Moreover, every node of organisation's network represents knowledge worker with individual set of knowledge and own cognitive and social potentials for learning (self-learning) and teaching (training). The knowledge worker, in every step of simulation, is looking for best source of knowledge. In addition, depending of node's neighbourhood the knowledge can be forgotten.

The existing diffusion models from literature were not suitable for competence modelling due to their limitations. The most important drawback is the lack of simultaneous support of vertical and horizontal diffusion. Moreover, diffusion logic proposed in literature does not reflect the competence development process. In our approach the diffusion logic is set to search for best teacher (source of knowledge) in node's neighbourhood. The best teacher is a node with the highest value of knowledge and teaching ability. The diffusion result is affected by the learning/teaching abilities of nodes, initial value of knowledge, vertical diffusion form other layers (relation between knowledge) and forgetting process. Similarly constructed diffusion model cannot be found in the literature.

The rest of the article is organized in the following way. Next chapter covers the issue of competence development in organization. We listed components associated with competence and developed a way to include them in the diffusion model. After that, the knowledge diffusion models are analysed. We concluded that all of them are lacking some proposition for competence development. Chapter four is dedicated to the description of multilayer knowledge diffusion model for competence development in an organization. The model is the focal point of the article. The last chapter is dedicated to validation of proposed model. In this chapter some case studies will be presented.

## 2  Competence development in an organization

### 2.1  Competence-based approach to system design

In literature we can find different definitions of competence linked by three fundamental characteristics: resources, context, and objectives [10]. The competence profile is data about a competence that may be aggregated for communication among individuals, organizations, and public administrations. The competence modelling issue has been a subject of research for a long time, starting with Frederick Taylor [11]. However, in recent years studies have greatly accelerated. An interesting review of competence notion can be found in [8,12,13]. The history and background of standardization in this area and research project are covered in [7]. Some computational approach for competence profile processing are described in [14,15]. The fuzzy nature of competence description is explained in [16]. Moreover, there is a number of high quality scientific journals with special issue dedicated only to competence including: Competence Management in Industrial Processes [17], Skills Management – Managing Competencies in the Knowledge-based Economy [18], Learning Networks for Lifelong Competence Development [19], Assessment of Competencies [20], Competencies Management [21].

Competence-based approaches have proved to be a critical tool in many organizational functions, such as employment planning, recruitment, trainings, raising work efficiency, personal development, managing key competencies [22]. In addition, competence-based system can be used for different purposes such as staff development and deployment, job analysis or economic evaluation [23]. The reasons for this are made by [24] as:

- Competence-based approach can provide identification of the skills, knowledge, behaviours and capabilities needed to meet current and future personnel selection needs, in alignment with the differentiations in strategies and organizational priorities.
- Competence-based approach can focus the individual and group development plans to eliminate the gap between the competencies requested by a project, job role, or enterprise strategy and those available.

The important way of competence developing is a community of practice, because growing number of people and organizations in various sectors are now focusing on communities of practice as a key to improving their performance [10]. Communities of practice are groups of people who share a concern or a passion for something they do and learn how to do it better as they interact regularly [25].

From a pragmatic point of view competence is a combination of components, usually related to knowledge, experiences and skills/abilities. It is important to notice that it is not possible to directly develop another person's competence. It is just possible to set the scene, to provide the tools and act like a catalyst [26]. As a result, the competence development is regarded as the acquisition of a specific set of competence's components (e.g. knowledge, skills) that constitutes a particular competence [9]. Moreover, overriding principle for development of competence becomes transmitting such attributes (components) to those people who do not possess them by range of activities, such as general communication, classroom teaching, on-the-job training and job rotation. [9]. The data about the competencies value/state is produced and transformed by identification, assessment and acquisition processes [27]. Competencies can be processed because there is a certain set of tools used to test competencies and estimate their levels [22,28].

There are some challenging issues with competence-based approach [29]: development and use of a consistent set of concepts and vocabulary for describing competences; classification of the different kinds and levels of activities within organizations that collectively contribute to achieving competence; articulating the interactions of different kinds and levels of organizational activities that are critical in processes of competence building and leveraging.

In order to overcome this challenging issue following attributes had to be identified and analysed [30]: how roles are assigned to employees; guiding principles; defined organizational processes; organizational culture (including values, atmosphere, and practices); organizational knowledge; managerial practices; organizational learning; information and information technology systems; work environment.

From observation of the organization we can observe that in any organization the competence are considered at following levels [2]: individual competence: competence of a person; collective/team competence: competence emerging from a group of persons; global/organizational competence – describes organizational ability of an enterprise.

The literature proposed some content of particular levels [30,31]:
- Individual competence (e.g. result orientation, role commitment, continuous learning, networking, creativity, intelligence, behavioral traits (including such aspects as honesty and maturity) ,motivation, communication capabilities);
- Team competence (e.g. knowledge sharing, cultural integration, resources utilization, innovation, management/ leadership);
- Organizational competence (e.g. knowledge landscape, knowledge assets, information sharing, push/pull power balance, synergy creation).

The dichotomy between definitions of competence that target individual workers and definitions that target the results of their work is a complex issue [32]. On the one hand the literature has focused on individual competences and has taken the worker's attributes as a starting point for discussing competence [33]. The worker's competence value is treated as a stock that can be developed through training and validated in "objective" rating schedules [34]. On the other hand, the competence is conceptualised as a characteristic of organisations where human competences are seen as one of the resources available to organisations [33].

## 2.2 Competence as a union of associated components

For the propose of model design it is required to identified the structure of competence. Competencies are considered as an union of different components (see Tab. 1). Thanks to literature analysis (e.g. [7,29,35]) we can distinguish some base components like: knowledge, skills, experience, etc. However, the competencies are not themselves resources in the sense of knowing how to act, knowing how to do, or attitudes [36]. The goal of competencies is to mobilize, integrate and orchestrate such resources. It is important to notice that all components presented in Tab.1 can be somehow measured.

**Tab.1: List of main competence components**

| Competence components (categories of resources) | Base references |
|---|---|
| − Knowledge,<br>− Skills,<br>− Abilities<br>− Behaviours | Treasury Board of Canada Secretariat (http://www.tbs-sct.gc.ca/) |
| − Knowledge,<br>− Skill,<br>− Ability<br>− and/or other deployment-related characteristic (e.g. attitude, behaviour, physical ability) | HR-XML (www.hr-xml.org)] |
| − Input Competencies<br>   o Knowledge<br>   o Skills<br>− Personal Competencies<br>   o Core Personality Characteristics<br>− Output Competencies | [37] |
|     o   Demonstrable Performance | |
| – Know-What<br>– Know-How.<br>– Know-Who (how closely one is acquainted with someone, knowledge from social network) | [38] |
| – Knowledge<br>– Know-how<br>– Behaviours | [39] |
| – Knowledge (what you learn in education)<br>– Experience (what you gather in your job, at your workplace and in social life)<br>– Abilities (to use your knowledge and experience) | [26] |
| – Knowledge (Theoretical knowledge, Knowledge of the existing, Knowledge of procedures<br>– Know-how (Procedural know-how, Empirical know-how)<br>– Know-whom (Relationship aspects, Cognitive capacities, Behaviours) | [39] |
| – Knowledge (includes theoretical knowledge and procedural knowledge)<br>– Skills (includes formalized know-how and empirical know-how<br>– Behavioural aptitudes (individual's behaviour at work) | [23] |

After the presentation of the components of competence it is crucial to understand what the relationship between them is. A good way of understanding the relationship is use of competence ontology structures, which can be found in literature [14,24,40,41,42]. Most of them is designed to [41]:

– define an organization-wide role structure based on the competencies required by job functions and organizational positions;
– identification of the competencies required in order to perform the various activities involved in each business process and assignment of roles to process activities based on these competencies;
– identification of the competencies acquired in the organization and assignment of users to roles through competence matching.

Moreover, the competence ontology is the most important part of an effective competence management system [24]. The competence's ontology is important because competence management system has to collaborate with other similar systems or e-learning and human resources applications. More formal approach to competence ontology building, based on the Description Logics, can be found in [42].

## 3 Models of knowledge diffusion

### 3.1 The nature of the problem

Diffusion of knowledge can be analysed in several dimensions. Knowledge diffusion, treated as a part of an innovative process, is the process by which an innovation is communicated through certain channels over time among the members of a social system [43]. Research shows the importance of social network structure, which should be analysed, developed and managed for continuous innovation in organizations [44].

In the area of scientific research knowledge diffusion can be defined as the adaptations and applications of knowledge documented in scientific publications and patents [45].

Technology diffusion is a complex social communication process. According to [46] technology diffusion starts with an innovation generated by a particular source. After that, potential adopters are informed about the availability of the new technology and persuaded by contact with prior users to adopt them. It is important to understand that technology diffusion is very sectoral. Ribeiro et al. [47] shows that any given innovation related to an area X is intensely used and diffused only within its specific area and will hardly permeate into other areas.

In a learning organisation knowledge diffusion is a process of knowledge communication and learning [48]. Moreover, the close relation among members results in strong willingness of knowledge diffusion.

It is important to say that tracing knowledge diffusion is a challenging issue due to the extreme complexity of diffusion processes [45]. Morone and Taylor [49] point out that knowledge diffusion is a complex social phenomenon which consists of, among others, knowledge spillover, knowledge transfer and knowledge integration. The nature of human interactions and information flow is affected mainly by the creation of new knowledge and the process of learning at an individual level [50]. Moreover, the different organizational and teamwork structure conducts different knowledge behaviours and their performance. However, we should have in mind that for group of people learning efficiency is in most cases accelerated [51]. In addition, diffusion makes neighbouring agents tend to display similar knowledge levels [52]. Social influence theories provide an interpretation that different social proximity evoke distinguish contagion effects [51]. The best learning outcome can be determined by the best suit payoff schemes and network structure changes within a complex social network [50]. In other words the effectiveness of the diffusion is a function of the network structure and seeding strategy used in delivering the initial broadcast [53].

From the economic point of view the knowledge diffusion process is related to the transfer of intellectual capital. Knowledge diffusion takes place through worker mobility [54] and the research task is related to finding the equilibrium between the host and the mobile worker. The [55] model offers a quantitative approach to explore the dynamic relationship between knowledge value and enterprise benefits in a given period.

At cognitive level the research of knowledge diffusion is related to the problem of [56]: how do individuals perceive and cognitively represent the social networks that surround them, and how do individuals' perceptions of their social networks affect their behaviours and outcomes?

The next important issue, related to knowledge transfer, is homophily, defining as tendency of people to associate relatively more with those who are similar to them than with those who are not [57]. The Golub and Jackson [57] show that homophily and the segregation induces in networks has important consequences for processes of interest, particularly the ones of information flow.

Knowledge diffusion must be based on efficient communication channels between all actors. The importance of such efficient channels is empirically supported by McGarvie [58] who shows that technological knowledge diffusion is faster in countries which share a common official language, whose inventors communicate more frequently by phone, and are geographically closer to each other.

The process of diffusion of knowledge is based on several communication mechanisms [59]: formal way of communication through documents, databases, face-to-face meetings, e-mails, videoconferencing, and social communication (excluding commercial transaction) throughout communities-or networks-of practice. Typical artefacts are opinion, practice, know-how [60]. The knowledge benefits can be externalized from the following three knowledge sources [61]: (1) the use of original knowledge inside the organization; (2)

the improvement of original knowledge due to internal investment; and (3) the integration of innovative knowledge.

From one point of view knowledge diffusion is intended by the organization. According to Canals [59] the diffusion process takes place in a formal way through the use of documents and databases or through interaction in face-to-face meetings or by using technological means as e-mail or videoconference. Form the another point of view unintended diffusion of knowledge is performed in accordance with knowledge spillovers process. The diffusion process takes advantage of the social relationships between employees of the firms, be it of a professional type through communities-or networks-of practice or more of a personal nature [59]. The most important issue is to combine in knowledge diffusion the intended knowledge diffusion mechanisms and unintended spreading of knowledge. For such reason knowledge diffusion is not equivalent to other diffusion processes modelled in natural sciences as epidemics or in social sciences (e.g. like the spread of rumors) [59]. Nevertheless, there are some attempts to do that e.g. tacit knowledge diffusion modelled as a SIR epidemic transformation [62].

### 3.2 Knowledge diffusion modelling

The problem of knowledge diffusion is an important element of complex network theory application. Based on the literature analysis we can recognise two approaches to problem modelling [63]. The first one is focuses on knowledge exchange behavioural patterns between a pair of individuals. The cognitive and social psychology and economics investigated absorptive capability, effectiveness and stimulation of knowledge share [64]. The second one used computer simulation to discuss the influence of the topology of social networks [63]. The simulation results show in what way the network structural characteristics influence knowledge diffusion.

When discussing the knowledge diffusion modelling we should keep in mind two dimensions of this problem: network topology and design of interaction rules driving knowledge transmission. Many studies show that the effectiveness of the diffusion mainly depends on the network structure and the seeding strategy used [65]. The problem of network topology analysis is solved by the utilisation of an existing network models which support real-world phenomena such as power-law ("scale-free") degree distributions, high clustering, short network diameter. In addition some authors make a debate about how accurately present models and corresponding analytic solutions or simulations render real-world network [66].

The main concept of knowledge transmission mechanism, according to [66], are: pay-off based models [67], opinions vectors, continuous or discrete, one-dimensional [68] or n-dimensional [69]. The other issue is a mechanisms of knowledge diffusion, most of models is aiming to maximizing the spread of influence in a network and they are based on assumed rather than measured influence effects [70].

The complex nature of knowledge diffusion problem is difficult to conceptualise and formalize. However, there are some propositions in literature. In [71] formal approach to create integrated ontology, which covers a number of learning activities, is proposed. Due to utilization of OpenCyc framework the way to computational semantics is clear. The real application of complex ontology, which is formalised for Computer-Aided Control Engineering, can be found in [72]. However, before we begin to think of the formalization we should try to define the based learning process in order to recognise the objectives outside the individual, and the transformation of these activities into measurable, efficient behaviour [73]. The presented literature gives some idea how to formalized different parts of knowledge diffusion model.

### 3.3 Related work in the areas of knowledge diffusion models

A number of papers studied a model of a population of agents whose interaction network co-evolves with knowledge diffusion and accumulation. General idea of the proposed model is based on the Cowan and Jonard model (CJ) [74]. Similar to CJ model the proposed model is designed to capture effects of incremental innovation and their diffusion over a network of heterogeneous agents. The CJ model assumes [75]:

- agents are arranged in one-dimensional space;
- each agent occupies one vertex and may interact with their $k$ nearest neighbours on either side;
- the population of individuals is endowed with different levels of initial knowledge;
- a small number of agents are 'expert' and are endowed with a high level of knowledge in at least one value of the vector;
- all individuals interact among themselves, exchanging information;
- knowledge is a non-rival good and can be transferred without decreasing the level of knowledge possessed by each trader.

In our work we extended classical CJ model to multidimensional vertical and horizontal diffusion scheme. Moreover, new mechanism of knowledge processing was introduced includes self-learning and forgetting process. Some authors noticed the importance of this factors (e.g. dissemination ability and knowledge forgetting in Xiaoqing and Runqing work [76]). However, this phenomenon is not regularly analysed because of the complexity.

In [77] two processes on the network are proposed: knowledge diffusion refers to the distribution of existing knowledge in the network, while knowledge upgrade means the discovery of new knowledge. Additionally, authors took into account the agent's knowledge absorptivity and forgetting factors represents some cognitive ability of agent. However, the proposed model works only for one type of knowledge, absorptivity and forgetting factors are constant and not associated with agent's network localisation.

The paper [78] focuses on knowledge diffusion as an economic process of different types of knowledge exchanging. Similar to previous work the paper covers the knowledge diffusion process (agent broadcasts his knowledge to the agents to whom he is directly connected) and knowledge creation process (agents receive new knowledge which is combined with their existing knowledge stocks). However, this paper only examines the relationship between network architecture and aggregate knowledge levels.

The key factors that affect the speed and the distribution of knowledge diffusion are identified in Morone et al. [79]. During knowledge transfer, knowledge is mastered through a backward process by which it is confronted and articulated with previous knowledge represented in form of a cognitive map. This paper is focused on knowledge dimension during diffusion process and do not cover the issue of agent's capabilities and diffusion of different kind of related knowledge.

The diffusion of different kind of knowledge in an organisation can be interpreted as a multilayers network (share the same set of nodes connected with many links grounded on different layers). In literature the main emphasis is placed on multi-layered diffusion processes through a multi-layered material for a wide range of applications, including industrial, biological, electrical, and environmental areas [80]. Nevertheless, there is some activity in area of multi-layered knowledge diffusion. The example is the work [81], where numerical simulation described non-linear relation between layers represents various network models. Diffusion model are the linear threshold and independent cascade. There are only focused on diffusion process excluding knowledge upgrade/creation process. Moreover, the model missed the relation with cognitive characteristic of actors and related self-learning and forgetting processes.

So far there has been no work combining the competence with knowledge diffusion. Moreover, the proposed model with vertical and horizontal diffusion for multilayer organisation graph with self-learning and forgetting processes is a new approach to the problem of knowledge diffusion.

## 4 Model of multilayer knowledge diffusion

### 4.1 Approach to competence modelling in the diffusion model

Modelling of the competence development based on the knowledge diffusion process requires a new approach. They are a couple of reasons for that. Competence cannot be changed directly, we can only influence its components. The value of the components of competence (see Tab. 1) can be determined.

In the proposed model, the components of competence are represented by the layers in the multilayer graph. From the point of view of the diffusion process, the content of layers is invalid and the diffusion process is focused on knowledge flow and not their content or meaning. We define only the relationship between their elements, which may be damping (weakening) or forcing (strengthening). This approach is similar to Shannon's Information Theory where content of the messages has no meaning. However, there is an opinion in the competence literature that the process of competence computing should be understood as enabling the use of competence databases for inference and combination of competencies for different functions and processes, not as a reductionist account of competencies to numeric models [14]. In our case we focused on knowledge flow and linked it to the diffusion process, which is based on mathematical transformations.

Proposition of 3th class' layers model for diffusion model for competence development [29]:
- Class 1: know-how—practical, hands-on forms of knowledge gained through incremental improvements to products and processes.
- Class 2: know-why—theoretical forms of understanding that enable the creation of new kinds of products and processes.
- Class 3: know-what— a strategic form of understanding about the value creating purposes to which available know-how and know-why forms of knowledge may be applied

We assume that one layer in our model is dedicated to one kind of knowledge, which belongs to one class of competence's components . The question arise: Can each component of competence be called knowledge? The works [10,25] prove that it is correct and knowledge can be expressed in various forms:

- Knowledge can be explicitly formalized – texts, documents multimedia.
- Knowledge can be a practice – it rests on the accumulation of experiments.
- Knowledge can be tacit – all cannot be formalized. Its transmission requires suitable means: conversation, training, joint work, etc.
- Knowledge can be social – the technical know-how of a company does not rest on an individual but on the interaction of all the members of its technical community. It is while collaborating, by confronting their points of view, that these technicians create and finally hold new knowledge.
- Knowledge is dynamic, and evolves/moves in time.

In our approach the term *knowledge* will be used for description of all components of competence. Based on the term *knowledge* all competence's components like instances of class: knowledge, skills, behaviour can be modelled in a multilayer graph as a single layer.

### 4.2 Knowledge network definition

#### 4.2.1 Knowledge domain

Every node in the network represents single knowledge worker. According to [4] the knowledge employee's main tasks related to knowledge are capture/extract, analyse/organise, find, create/synthetize and distribute/share. In the organisation's network different types of knowledge is propagated in order to acquire competence by employee. Moreover, the productivity of knowledge workers is enhanced through competence enhancement and learning which take place directly at workers' workplaces [82].

*Definition: Organization members*

The organization $X$ composed of knowledge worker determined by index $i$. $I = \{i : i \in N_+\}$

*Definition: Knowledge domain in organisation*

*The knowledge domain in organisation $X$ is represented by the set $K = \{k_j : k_j \geq 0\}$, the elements of set are indexed by $j$, $J = \{j : j \in N^+\}$. The element $k_j$ is a knowledge element that represents some related part of domain of discourse. In the analysed domain the last position of $J$ is $j*$.*

The methods of knowledge modelling mainly focus on the formally representation of relationship between different areas/element/types of knowledge. The best way to do is to use the ontological approach. Ontology is a formal, explicit specification of a shared conceptualization [83]. The main components of an ontology are concepts, relations, instances and axioms [84]. The relations describe the interactions between concepts or a concept's properties. In the problem of knowledge diffusion is essential to show the existence of the relationship between areas/elements of knowledge. Because defining the types of relationship between is not very useful for numerical processing and difficult to determine, we focused on causality and mutual order of knowledge areas/elements. Therefore, we choose the method of presentation of the knowledge domain based on the Knowledge Space Theory [85].

In our approach the competence acquisition is a result of combination of different competence's components [86]. In other words the proper combinations of different knowledge elements, which reflect competence's components, result in an efficient competence acquisition by a knowledge worker. Competence is an ability to find an effective way to theoretical knowledge usage in order to solve a practical problem and the ability to verify the solutions. We use the Knowledge Space Theory to describe the relation between knowledge element in domain $K$.

Based on the Knowledge Space Theory let $(K, \leq)$ be a partial ordered set. In this theory the prerequisite relationship is cover by the surmise relation [87] and function $\prec$ represents prerequisite relationship. Two knowledge items $a$ and $b$ are in surmise relation $a \prec b$ if, whenever a person has solved/maintained item $b$ correctly, we can surmise that this person is also able to solve item $a$ correctly [88]. According to [89] we say that for $k_a, k_b \in K$, $k_b$ covers $k_a$ denoted by $k_a \prec k_b$ if $k_a < k_b$ and for $\forall k_c \in K$ and $k_a \leq k_c < k_b$ implies $k_a = k_c$. Moreover, the graph $(K, \prec)$ is a Hasse digram for $K$. The important assumption for future discussion is that due to cognitive nature of problem and mutual relation between them, not all potential knowledge states have to be observed [90].

Let $k_a, k_b \in K$, we say that element $k_a$ is maximal element in $K$ if for $\forall k_b \in K, k_b \geq k_a$ then $k_b = k_a$. We denoted this by symbol $k^{\max}$. Similarly, we say that element $k_a$ is minimal element in $K$ if for $\forall k_b \in K, k_a \geq k_b$ then $k_a = k_b$ and the dedicated symbol is $k^{\min}$.

For $(K, \leq)$ and $k_a, k_b \in K$, the lower shadow of $k_a$ is a set $\{k_b \in K, k_b \prec k_a\}$ denoted by $\Delta k_a$ and consequently the upper shadow of $k_a$ is a set $\{k_b \in K, k_a \prec k_b\}$ denoted by $\nabla k_a$. Moreover, for $k_a > 0$ all elements of its lower shadow $\Delta k_a$ have to be greater than zero $\{k_b : k_b \in K, k_b \prec k_a, k_b > 0\}$.

### 4.2.2 Knowledge network

In a knowledge network the node actively processes knowledge and edges represents channels for knowledge relocation [91].

*Definition: Knowledge network*

The knowledge network for organisation $X$ is $G^X = \{G_j\}$ multilayer graph of graphs $G_j$, where $j \in J$ is an index of knowledge in $K$. Every knowledge element form set $K$ is represent by a single layer in $G^X$.

*Definition: Knowledge layer*

Every layer in $G^X$ is a undirected graph without self-loops $G_j = (V, E_j, f_j)$. Where $V = \{v_i\}$ is a set of nodes representing knowledge worker $i$, $E_j \subseteq V \times V$ - set of edges represents symmetrical relationship between nodes (knowledge employee) on layer $j$, and $f_j : E_j \to \Re^+ \cup \{0\}$ is a variable edge labelling function.

For any node from $G_j$ we can recognise knowledge worker's neighbourhood. The $e^j(v_{i_1}, v_{i_2})$ is a binary variable for $v_{i_1}, v_{i_2} \in V$, and $i_1, i_2 \in I$. If connection between $v_{i_1}$ and $v_{i_2}$ exists at layer $j$ then $e^j(v_{i_1}, v_{i_2}) = 1$, otherwise $e^j(v_{i_1}, v_{i_2}) = 0$. The neighbourhood of node $i_1$ at layer $j$ is a set $\Gamma_{i_1}^j = \{v_{i_2} : v_{i_2} \in V, e^j(v_{i_1}, v_{i_2}) = 1\}$.

*Definition: Knowledge vector for worker i*

*Based on the multilayer graph $G^X$ we can formulate the knowledge vector for worker $i$:* $\overline{K}_i = [k_{1,i}, k_{2,i},...,k_{j,i},...,k_{j*,i}]^T$. *Where* $k_{j,i} \geq 0$ *and* $k_{j,i}$ *represent the value of knowledge on layer $j$ for worker $i$. Moreover, any worker $i$ has the knowledge set related in its own way:* $(K_i, \leq)$.

In order to distinguish personal abilities for communication on knowledge level two parameters are introduces:

1. Cognitive abilities for node $v_i$: $o_i \in <0,1>$ variable for cognitive abilities for node $v_i$. The highest $o_i$ the fastest actor behind $i$ is able to learn and acquisition knowledge form others in order to increase his/her knowledge level.

2. Social abilities for node $v_i$: $l_i \in <0,1>$ variable for social abilities for node $v_i$. The highest $l_i$ the fastest actor behind $i$ is able to teach others. This means that it has a social skills to adapt (personalize) communication to the recipient [92].

### 4.3 Structure of competence use in the model

Competence can get gradually stronger, in a situation where surroundings affect and stimulate its components. For example, we acquire new skills in a training session or while working (e.g. software developers programming everyday). Competence (its level) can also degrade. The most common reason for it, is not using the given competence in everyday work. The other is thanks to technology progress which makes the components of competence outdated. We can distinguish different relations between competencies which affect the interaction between them. Increasing competence in a certain competence group (e.g. communication) can affect the increase of other competencies (e.g. sales of products). Next issues regarding competence processing in an organisation start to show up when we take a look from a company's perspective. From the company's point of view, certain competencies are created only by combining the competencies of a greater number of employees. The complexity of these combined competencies is too great for a single person to obtain this kind of competence.

In this approach we do not analyse the content of knowledge included in competence. In our case we are interested in the competence's level, which allow us to analyse the knowledge and competencies growth and dynamics in the organization. In presented method the competence value will be normalized to range $<0,1>$ in order to be compatible with scale in literature. The level of competence is related to the expertise of an employee (see Tab. 2) [93]. According to cognitive science employees with more competence (expert) within their domains are skilled, competent and think in qualitatively different ways than novices do [94].

**Tab. 2: Linking competence value with expertise level**

| Approximate value of $c_i$ | Expertise level | Description (based on [95]) |
|---|---|---|
| $0-0,2$ | Novice | Minimal exposure to the domain. |
| $0,2-0,4$ | Initiate | Began introductory instruction to the domain. |
| $0,4-0,6$ | Apprentice | Undergoing a program of domain learning beyond the introductory level. Traditionally the apprentice is immersed in the domain by a more experienced employee. |
| $0,6-0,8$ | Journeyman | Person who can perform a day's labor unsupervised, although working under orders. |
| $0,8-1$ | Expert | The distinguished or brilliant journeyman, highly regarded by peers, whose judgments are uncommonly accurate and reliable, whose performance shows consummate skill and economy of effort, and who can deal effectively with rare or ''tough'' cases. Also and expert is one who has special skills or knowledge derived from extensive experience with subdomains. |
| 1 | Master | Master is any journeyman or expert who is also qualified to teach those at a lower level. Traditionally, a master is one of an elite group of experts whose judgments set the regulations, standards or ideals. |

*Definition: Competence*

The competence set for organization $X$ is defined in the following way: $C^X = \{c_a : a = 1...A, c_a \in <0,1>\}$

There relationships between competence and knowledge in organization $X$ are represent by the matrix: $M^X[t] = \|m_{ja}[t]\|$, $m_{ja} : s(C,K) \to \Re^+ \cup \{0\}$, where $s(C,K)$ is a variable for competence $a$ and knowledge $j$ relation strength representation at time $t \in T$. The relationships in the matrix are continuously changing due to companies priority and change in the market. If $m_{ja} = 0$ then there is no relations between competence $a$ and knowledge $j$. At the figure 1 the graphical representation of matrix $M^X$ is presented.

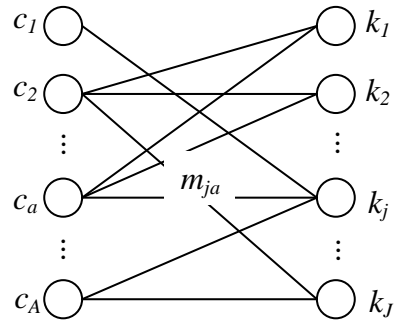

**Fig.1: Relation mash for competences and knowledge relation**

Based on the matrix $M^X[t]$ the competence value, at time $t \in T$, for worker $i$ is following:

$$c_a^i[t] = \sum_j m_{ja}[t] \cdot k_{ji}[t] \qquad (1)$$

The final value of $c_a^i[t]$ form (1) has to be normalized to fulfil $c_a^i[t] \in <0,1>$

## 4.4 Process definition

In order to analyse the competence development in an organization in addition to the structure of the network, which represents the relationships that exist between staff, we also need to describe the processes associated with the competence development. Let us introduce the time index: $T = 1,...,t,t+1,...,t*$.

### 4.4.1 Knowledge diffusion

In the proposed model the knowledge diffusion process takes place in two directions: vertical ($D_{j,i}^v$) and horizontal ($D_{j,i}^h$).

#### 4.4.1.1 Horizontal knowledge diffusion

Horizontal process of knowledge diffusion is related to knowledge diffusion between knowledge network nodes (representing employees) on a selected layer $j$. In fact, it involves simulating a situation in which the relationship will be created at the level of tacit knowledge. The relationship affects the knowledge of involved employees with regard to their ability to teach and learn. In the classic Nonaka's model this process is called socialization [96]. According to [97] for knowledge sharing to be most effective, it should take place between people who have a common knowledge and can work together effectively. The mutual relationship should be strong. Thus tacit knowledge sharing is connected to ideas of communities and collaboration.

Horizontal knowledge diffusion occurs only between active nodes. There are two possible methods of knowledge diffusion between nodes:
- Broadcast: the node transfers knowledge to all connected nodes;
- Multicast: the node transfers knowledge to a selected set of nodes. The set of receiving nodes may be selected random or based on some strategies.

In our approach we focus on the multicast scheme for horizontal knowledge diffusion. Every node on selected layer of multilayer graph $G^X$ is looking for most effective source of knowledge in its neighbourhood on this layer. In this context effective means with best combination of knowledge and social abilities. As a result the horizontal knowledge diffusion occurs only when the node is able to locate node with a greater potential for knowledge transfer in his neighbourhood. In other words weak looks for strong.

*Definition: Horizontal knowledge diffusion*

*Horizontal knowledge diffusion is calculated for node $i$ on layer $j$ based on the function $\alpha$ defined in following way: $D_{j,i}^h = \alpha(k_{j,z}, o_i, l_z, d_i, f_{j,z}) \to \Re^+$, where $k_{j,z}$ is a knowledge value for node $z \in I$ on layer $j$, $o_i$ is a cognitive (learning) abilities for node $v_i$, $l_z$ is an social (teaching) abilities for node $v_z$, $d_i$ is a component responsible for node $i$ degree distribution, $f_{j,z}$ is a strength of the relationship between nodes $j$ and $z$. The $z$ is a node $v_{j,z} \in \Gamma_i^j$ with maximal combination of cognitive and social abilities. The horizontal knowledge diffusion occurs only if $k_{j,z} \cdot l_z > k_{j,i} \cdot o_i$ for $v_{j,z} \in \Gamma_i^j$.*

The interpretation of knowledge diffusion function depends on the purpose and goals of the organization. In addition, the final function form depends on the specific nature and structure of knowledge networks and knowledge resources in an organization. For the purposes in the article we have proposed functions (2-10) for implementing the various described processes. It is clear that individual processes can be expressed by other functions depending on the intentions of the designer.

Let us propose some form of function $\alpha$:

$$D_{j,i}^{h}[t] = \frac{(k_{j,z}[t] - k_{j,i}[t]) \cdot l_z \cdot o_i}{d_z \cdot A} \cdot d_{\max} \cdot f_{j,z} \qquad (2)$$

where $d_{\max}$ is a maximal node degree on layer $j$ and node $z$ has $\max(k_{j,z} \cdot l_z)$ for $v_{j,z} \in \Gamma_i^j$. The $A$ is a fixed value. In function (2) we take under consideration the possible time to give of node $v_{j,z}$. If the node $v_{j,z}$ degree is higher than the selected knowledge diffusion is limited by other relations. Moreover, higher cognitive abilities and nodes relationship strength also support knowledge diffusion in positive way.

### 4.4.1.2 Vertical knowledge diffusion

Vertical knowledge diffusion takes place in a single node and occurs between its knowledge layers. Generally speaking the knowledge value increasing on layer $j$ may increase the knowledge value on other layers. The relationship between layers can be deducted from $(K, \leq)$ or described by organisation members and saved in an dedicated matrix. Moreover, the vertical knowledge diffusion process is an internal process in contrast to the horizontal knowledge diffusion which is an external process for a knowledge worker.

Let's define the vertical diffusion matrix for worker $i$: $M_i[t] = \left\| r_{j_1,j_2}^{i}[t] \right\|, M_i \subset K_i \times K_i$, where $r_{j_1,j_2}^{i} \geq 0$, $r_{j_1,j_1} = 0$, and $j_1, j_2 \in J$. The matrix may vary with time due to changing relation between knowledge according to technological change and innovation. The $r_{j_1,j_2}^{i}[t]$ described relations between layers and a single knowledge type at time $t$. If knowledge value $k_{j_1,i}$ is increasing then the value $k_{j_2,i}$ is increased according to $k_{j_2,i}[t+1] = k_{j_2,i}[t] + r_{j_1,g}^{i}[t] \cdot k_{j_1,i}[t]$.

*Definition: Vertical knowledge diffusion*

*Vertical knowledge diffusion for nodes $v_{\bullet,i}$ causes change of the knowledge value on layer $j \in J$ and affected other layers $g \in J \setminus \{j\}$ according to matrix $M_i$ at time $t$. Vertical* *knowledge diffusion is calculated for layers $g$ based on the function $\mu$: $D_{j,i}^{v}(k_{j,i}) = \mu(r_{j,g}^{i}, k_{g,i}) \to \Re$, where $r_{j,g}^{i}$ is a value of diffusion relationship between layers $j$ and $g$ for node $i$, and $k_{g,i}$ is a knowledge level for node $i$ on layer $g$. The function for vertical knowledge diffusion has argument $(k_{j,i})$ containing initial information about the value of knowledge increasing for node $i$ on layer $j$. This increase value has to be diffused between the other layers. The vertical knowledge diffusion occurs for node $v_{\bullet,i}$ if at any layer the knowledge value has been changed $\exists n \in J : k_{n,i}[t+1] \neq k_{n,i}[t]$.*

Let us propose some form function for vertical knowledge diffusion:

$$k_{j,n} = r_{j,n}^{i} \cdot k_{j,i} \qquad (3)$$

where $n \in J$ is a layer's index.

The function (3) is one of the possible linear relationships. For the real processes the matrix $M_i$ has to be defined by the expert from organisation with some cognitive competencies. As a result the relationship between different layers can be non-linear, nested and with feedback.

### 4.4.2 Knowledge deterioration (forgetting)

Over time employee competencies (knowledge) are reduced if they are not stimulated by the workers form surrounding and the work itself. From the formal point of view knowledge forgetting model can be found in [98]. The main concept is to incorporated in knowledge model fact that agents didn't always remember their previous knowledge (i.e. agents have perfect recall). Sometime we want to model the fact that certain knowledge is discarded. From the business point of view organizations must forget old habits in order to learn new and more appropriate ways of doing things [99]. Organisation may be forgetting knowledge intentionally (avoiding bad habits, unlearning) and accidentally (failure to capture, memory decay) [100]. From the cognitive science point of view, man develop their skills in an environment that stimulates them [94]. In this case, when my co-workers are less competent, with time my capacity will decrease (equal to their average level). A lot of works in psychology show that the environment impact on our activity. In the case of knowledge processes uninspiring surroundings causes progressively lose our knowledge.

Let us introduced a formula for average knowledge transfer potential for node neighbourhood calculation:

$$\tilde{k}_{j,i}[t] = \sum_{k_{j,q} \in \Gamma_i^j} \frac{k_{j,q}[t] \cdot l_q}{card(\Gamma_i^j)} \tag{4}$$

where $card(\Gamma_i^j)$ is a number of nodes in $i$ neighbourhood. If the value of (4) is less than worker knowledge acquisition potential (product of worker's knowledge level and his/her cognitive abilities) then the worker's knowledge starts to deteriorate.

*Definition: Knowledge deterioration (forgetting) process*

*If for node $i$ on layer $j$ following condition occurs $k_{j,i}[t] \cdot o_i \geq \tilde{k}_{j,i}[t]$ then forgetting process is described by the formula $F_{j,i} = \chi(k_{j,i}, \tilde{k}_{j,i}, o_i)$, where $\chi$ is a forgetting function $\chi(k_{j,i}, \tilde{k}_{j,i}, o_i) < k_{j,i}$, $k_{j,i}$ is a knowledge level for node $i$ on layer $j$, $\tilde{k}_{j,i}$ is an average knowledge transfer potential for node $i$ on layer $j$ neighbourhood, and $o_i$ is a cognitive abilities for node $v_i$. Moreover, the forgetting process occurs also vertically $D_{j,i}^v(-(F_{j,i}))$.*

In proposed function $F_{j,i}$ for node $i$ on layer $j$ the forgetting factor is related to node neighbourhood $\Gamma_i^j$ and knowledge worker's cognitive ability . In general, worker with high cognitive ability forgets slower and the worker is forced to start forgetting by the weakness of his neighbourhood.

*Assumption: non-zero knowledge condition*

*If the knowledge value for node $i$ on layer $j$ is greater than zero in the next steps of the knowledge value has to be always greater than $\Omega > 0$: $k_{j,i}[t] > 0 \to k_{j,i}[t+\hbar] \geq \Omega : \hbar = (1,...,t*-\hbar)$. The minimal value of knowledge value is represent by the variable: $\Omega > 0$*

One of the forgetting formula proposition is following:

$$k_{j,i}[t+1] = k_{j,i}[t] - \Xi_j = k_{j,i}[t] - \frac{(k_{j,i}[t] - \tilde{k}_{j,i}[t]) \cdot (1 - o_i)}{B} \tag{5}$$

$$D_{j,i}^v(-k_{j,i}) \to$$

$$\forall n \in J : k_{n,i}[t+1] = k_{n,i}[t] - \Xi_n = k_{n,i}[t] - r_{j,n}^i \cdot \left( \frac{(k_{j,i}[t] - \tilde{k}_{j,i}[t]) \cdot (1 - o_i)}{B} \right) \tag{6}$$

where $B$ is a fixed value. Moreover if for (5) $k_{j,i}[t] < \Xi_j$ then we set the value of $k_{j,i}[t+1] = \Omega$ and respectively for (6) the for $\forall n \in J : k_{n,i}[t] < \Xi_n \rightarrow k_{n,i}[t+1] = \Omega$. The variable $\Omega$ are the set by the organisation management lowest acceptable value of knowledge in system.

### 4.4.3 Self-learning

Due to the rapid obsolescence of knowledge and the requirements of increasingly complex processes there is a need to continuously acquire new knowledge by employees. Lifelong learning philosophy [101] assumes that any worker maintains ongoing, voluntary, and self-motivated pursuit of knowledge. In the proposed model this movement is described as a self-learning process.

*Definition: Self-learning process*

*If for node $i$ on layer $j$ following condition occurs $k_{j,i}[t] \cdot o_i < \tilde{k}_{j,i}[t]$ then self-learning process is described by the formula $S_{j,i} = \delta(k_{j,i}, \tilde{k}_{j,i}, o_i, cc_i)$, where $\delta$ is a self-learning function, $k_{j,i}$ is a knowledge level for node $i$ on layer $j$, $\tilde{k}_{j,i}$ is an average knowledge transfer potential for node neighbourhood, $cc_i$ is a clustering coefficient of node $v_{j,i}$, and $o_i$ is a cognitive abilities for node $v_i$. Moreover, the self-learning process occurs also vertically $D_{j,i}^v(S_{j,i})$.*

The function $\delta$ incorporated node's surrounding and cognitive ability into self-learning process. If the average knowledge level of $\Gamma_i^j$ is higher than node's knowledge then knowledge worker has to invest some time in order not to stand out from the rest and be valuable for communication. Moreover, the high clustering coefficient reflects larger environment that may have more pressure due to high cliqueness. In the article we propose following self-learning formula:

For $cc_i > 0$:

$$k_{j,i}[t+1] = k_{j,i}[t] + \Psi_j = k_{j,i}[t] + \frac{(\tilde{k}_{j,i}[t] - k_{j,i}[t]) \cdot o_i \cdot cc_i}{C} \quad (7)$$

$$D_{j,i}^v(k_{j,i}) \rightarrow$$

$$\forall n \in J : k_{n,i}[t+1] = k_{n,i}[t] + \Psi_n = k_{n,i}[t] + r_{j,n}^i \cdot \frac{(\tilde{k}_{j,i}[t] - k_{j,i}[t]) \cdot o_i \cdot cc_i}{C}$$

$$(8)$$

For $cc_i = 0$:

$$k_{j,i}[t+1] = k_{j,i}[t] + \Psi_j = k_{j,i}[t] + \frac{(\tilde{k}_{j,i}[t] - k_{j,i}[t]) \cdot o_i}{D} \quad (9)$$

$$D_{j,i}^v(k_{j,i}) \rightarrow$$

$$\forall n \in J : k_{n,i}[t+1] = k_{n,i}[t] + \Psi_n = k_{n,i}[t] + r_{j,n}^i \cdot \frac{(\tilde{k}_{j,i}[t] - k_{j,i}[t]) \cdot o_i}{D}$$

$$(10)$$

where $C$ and $D$ are fixed values.

## 4.5 Procedure of knowledge diffusion

Competence development based on the knowledge diffusion involves various processes: horizontal knowledge diffusion, vertical knowledge diffusion, knowledge forgetting, self-learning. In addition, knowledge diffusion is a two dimensional process. In this section we will develop the main points of procedure for knowledge diffusion calculation in multilayer networks.

Layer selection is based on layer's ranking $\Re^{layyers}$, calculated based on vertical diffusion matrix for variable $\hat{k}_w$:

$$\hat{k}_w = \sum_i \sum_g r_{w,g}^i \quad (11)$$

where $w, g \in J$. As a result the ranking consists of indexes of layers with order determined by the value of $\hat{k}_w$. Taking into consideration all worker we analysed what layer has the biggest influenced on other layers $\max(\hat{k}_w)$. If the layers have the same value of $\hat{k}_w$ the order is selected randomly. The ranking starts with layer with biggest $\hat{k}_w$.

Node selection is based on node's ranking $\mathfrak{R}_j^{nodes}$, generated with regards to knowledge level of all nodes on analysed layer $j \in J$. Due to 'pull' nature of described knowledge diffusion process the ranking starts with the node with lowest value of knowledge.

Procedure of knowledge diffusion:
FOR all subsequent element $l \in \mathfrak{R}^{layyers}$

    FOR all subsequent element $n \in \mathfrak{R}_l^{nodes}$

- Execute horizontal knowledge diffusion process: $D_{l,n}^h$
- Execute vertical knowledge diffusion process: $D_{l,n}^v(k_{j,i})$
- Calculation average knowledge transfer potential for node neighbourhood $\tilde{k}_{l,n}$
  - If $k_{l,n} \cdot o_n \geq \tilde{k}_{l,n}$ then execute forgetting process $F_{l,n}$
    - Execute vertical knowledge diffusion process: $D_{l,n}^v(-k_{j,i})$
  - If $k_{l,n} \cdot o_n < \tilde{k}_{l,n}$ then execute self-learning process $S_{l,n}$
    - Execute vertical knowledge diffusion process: $D_{l,n}^v(k_{j,i})$

The logic of horizontal diffusion refers to the analysis of components' values on single level. Therefore, may not matter what granularity scales are taken for each competence's component because all related operation taken place on selected layer. The problem could arise in terms of relations between layers (which represent competence components). Mutual vertical relations between layers can be nonlinear and based on own logic. In presented approach the matrix $M_i$ is dedicated to competence based transformation. If the competence components are reflected at different levels of granularity then we need to maintain the normalization process.

The presented approach, from the methodical point of view, is an agent based simulation. This kind of simulation is adequate to such systems with complex interaction. In our simulation we based on the NetLogo framework [102].

# 5      Areas of applications

## 5.1    Competence management paradigm

The model of knowledge diffusion is used to analyze the development of competencies in an organization. However, its final application is competence management. Keeping in mind main management axioms [103] we are going to discuss whether competence management is possible based on the proposed model:

- Axioms of management 1: The object is suitable for observations and measurements.
  The presented model gives ability to observe and measure every component of competence. Moreover, all knowledge flow between actors can be tracked and analysed.
- Axioms of management 2: At the interval of observation object can change its state.
  The dynamics of knowledge flows on different levels of network is noticeable. The process of knowledge diffusion is based on the continuously changing value of actors (workers) knowledge.
- Axioms of management 3: The predetermined target defined expected object's state.
  In the organization the target is set on the strategic level and concerns for expected values of competencies.
- Axioms of management 4: There are alternative ways to influence the behaviour of an object
  Any types of knowledge, components of competence, can be changed by training (increase of knowledge level ), expert's mentoring (direct diffusion of knowledge) or team building (network reconfiguration).
- Axioms of management 5: There is a pre-defined criterion of management efficiency.
  The criterion determines the degree of matching acquired competences to market or company requirements.
- Axioms of management 6: There are resources for the execution of the decision.
  The network consists of nodes which represents knowledge workers (actors).

Moreover, in the discussed context, competence management is a process of tracking changes in the content of knowledge related to the competences.

## 5.2 Simulation model

All the concepts of knowledge diffusion models require validation. In real conditions only few models can be checked due to the limitation of data. As a result, a number of simulation network models is used. The description of models can be found in [104]. The literature review shows that the diffusion models are validated based on the Watts-Strogatz model [105]. The Watts-Strogatz model reflected the "small-world" characteristic of complex network. According to Cowan and Jonard [74,78] "small world" networks generate the fastest knowledge growth. Moreover, Cowan and Jonard found that the steady-state level of average knowledge is maximal when the network structure is a small world, It means that most connections are local, but roughly 10% of them are long distance. The relatively big clustering coefficient is beneficial for knowledge diffusion in the agents' network [48]. The Watts-Strogatz model combines a strong degree of local cohesiveness with a small fraction of long distance links permitting knowledge to be circulated rapidly among distant parts of the network

At the beginning the network generated by the Watts-Strogatz model is a regular network, and it can rewire from the regular network to the random network by adjusting the parameter $p$. In the literature in order to generate a small world network for diffusion process validation the parameter $p$ is set $p = 0,1$ [74,78,79,106]. For $p = 0$ the network is regular and $p = 1$ generated random network.

The analyzed simulation model contains 500 nodes (agents). The network is generated based on the Watts-Strogatz model for $p = 0,1$. We are going to simulated $4^{th}$ layers model, where the relations between the layers are following: $n_2 < n_1, n_3 < n_2, n_4 < n_2$ (see Fig. 3). For a different simulation propose the value of the competence matrix $M^X$, the vertical diffusion matrix $M_i$ (the same for all workers), node's cognitive and social abilities, and knowledge vector for worker will be set randomly. The workers are divided into two groups, 'normal' workers and experts with knowledge level significantly greater than other workers.

## 5.3 Applications

Due to high stochastic nature of competence development process and multidimensional of the proposed model the deep simulation analysis is very difficult to maintain. In order to illustrate the different aspects of the proposed model, in the context of competence management, we will discuss a number of case studies.

### 5.3.1 Modelling the competence development based on multilayer diffusion

The proposed model was verified during simulations in terms of multilayer diffusion and development of competence. Simulations were performed on Wats-Strogatz network with 0.1 rewiring probability. Initial knowledge to each worker was assigned randomly from the range (0,5) and maximal expert knowledge was assigned to the level of 30. Number of experts was assigned to 3% of all network nodes. Simulations were performed at model parameters with assigned values A=2.0, B=0.1, C=2.0 and D=2.0 (for formulas 2-10). Structure of competence $c_1$ was based on the vector (0.5, 0.5, 0.0, 0.0) assigned to knowledge $k_1,k_2,k_3,k_4$ respectively, for $c_2$ (0.10, 0.20, 0.30, 0.40), for $c_3$ (0.25, 0.25, 0.25, 0.25) and uniform distribution for $c_4$ with values (0.40, 0.40, 0.10, 0.10). At the first stage simulations were performed without vertical diffusion and show process for knowledge diffusion independent for each component of competence. In the second stage settings for vertical diffusion were based on symmetric relation between layers with same symmetric intra layer diffusion at the level 0.4 with relations showed in Fig. 2. Third stage of simulation was configured using structure of dependencies and matrix setting showed respectively in Fig. 3 and Fig. 4.

|   | 1 | 2 | 3 | 4 |
|---|---|---|---|---|
| 1 | 0 | 0,4 | 0,4 | 0,4 |
| 2 | 0,4 | 0 | 0,4 | 0,4 |
| 3 | 0,4 | 0,4 | 0 | 0,4 |
| 4 | 0,4 | 0,4 | 0,4 | 0 |

**Fig. 2: Symmetric settings for vertical diffusion**

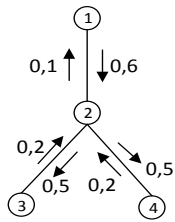

**Fig. 3: Hierarchical structure for knowledge**

|   | 1 | 2 | 3 | 4 |
|---|---|---|---|---|
| 1 | 0 | 0,6 | 0 | 0 |
| 2 | 0,1 | 0 | 0,5 | 0,5 |
| 3 | 0 | 0,2 | 0 | 0 |
| 4 | 0 | 0,2 | 0 | 0 |

**Fig. 4: Matrix based setting for vertical diffusion**

Results for model with disabled vertical diffusion is similar to earlier models are presented in Fig. 5. For each knowledge was activated learning, self-learning and deterioration parameter. Results based on assigned parameters show ability to simulate processes and changes within results in terms of deterioration and acquiring knowledge. In the next stage, the simulation was performed with active vertical diffusion based on symmetrical settings and the result is presented in Fig. 6.

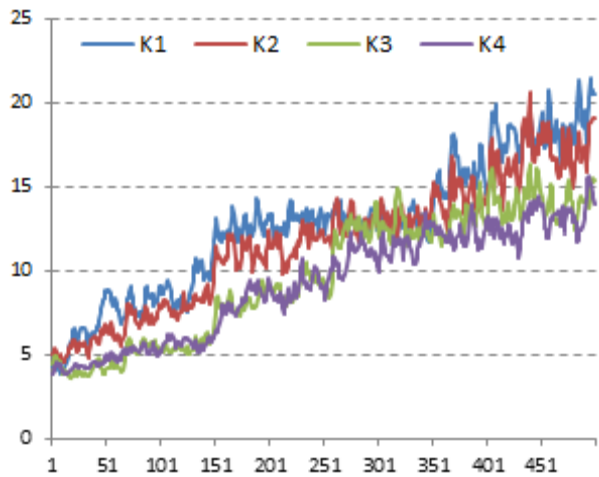
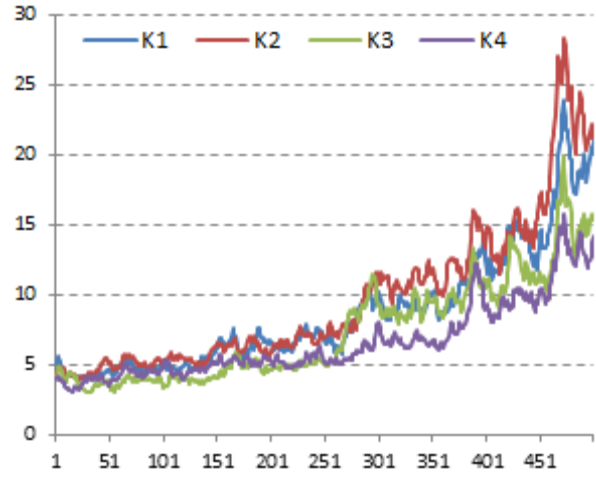

Fig. 5: Results of simulation with disabled vertical simulation

Fig. 6: Results of simulation with symmetric vertical simulation

Simulation model allows to track incoming knowledge incoming and outgoing for each component. Results for incoming and for outgoing knowledge at each layer are visible in Fig. 7 and Fig. 8. Monitoring knowledge diffusion in terms of incoming and outgoing knowledge makes possible to track effectiveness and implement strategies improving the process. Using presented approach it is possible to detect problems at this level and make diffusion more effective. Activated vertical diffusion resulted in higher total average knowledge in each layer obtained in the end of process.

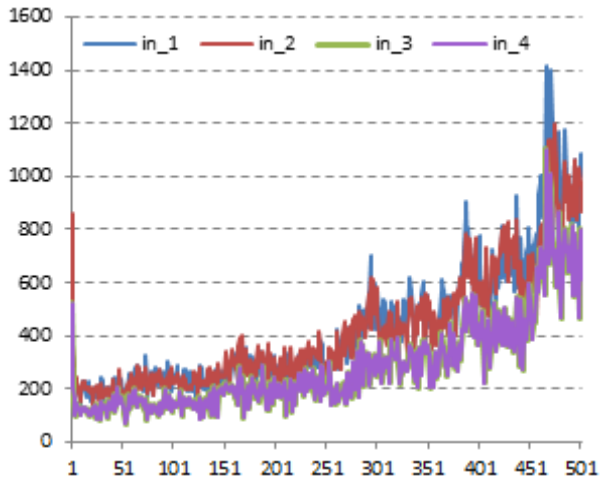
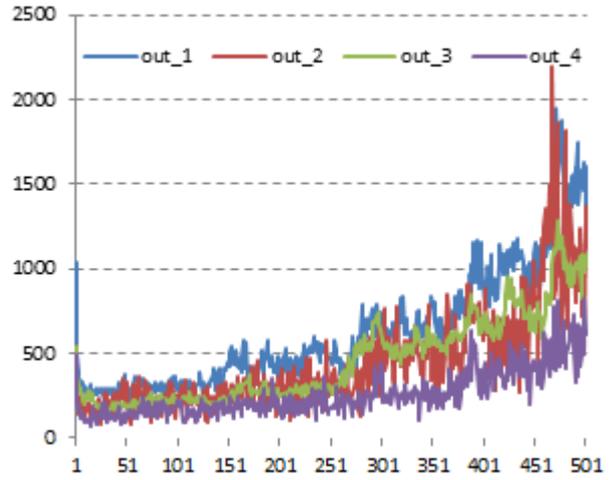

**Fig. 7: Knowledge incoming form vertical diffusion based on symmetric relations**

**Fig. 8: Knowledge outgoing from vertical diffusion based on symmetric relations**

Combining within model knowledge components and building metric for competences makes possible to track different competence development over time. In Fig. 9 competence development for each competence that can be modelled using different relations between knowledge components and each competence is shown.

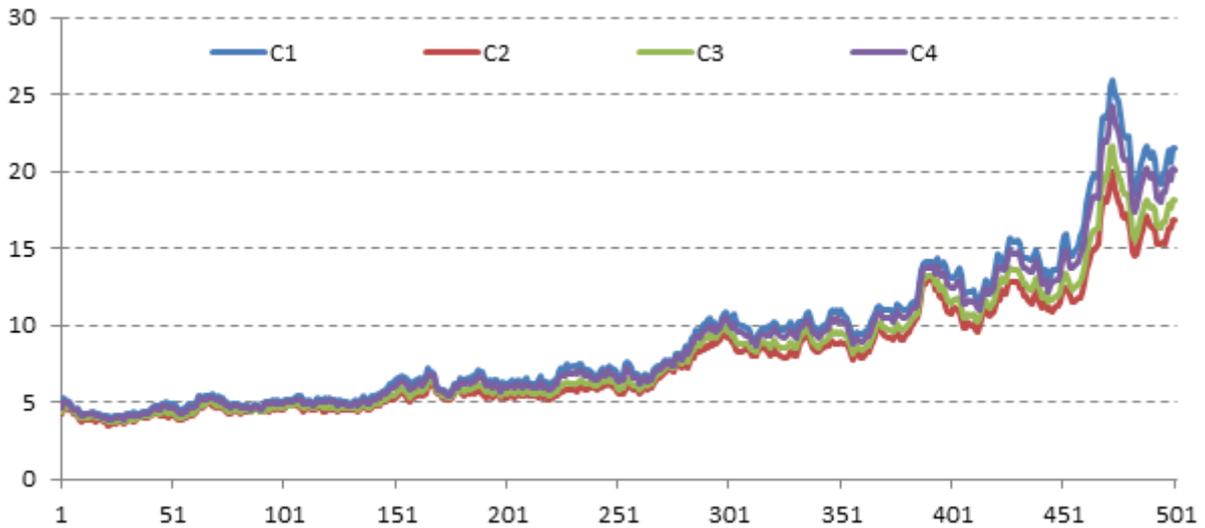

**Fig. 9: Development of competences based on four layers**

In the next step simulation is based on the asymmetric settings for vertical diffusion and results for knowledge in each layer are shown in Fig. 10. Results for incoming knowledge in each layer are visible in Fig. 11.

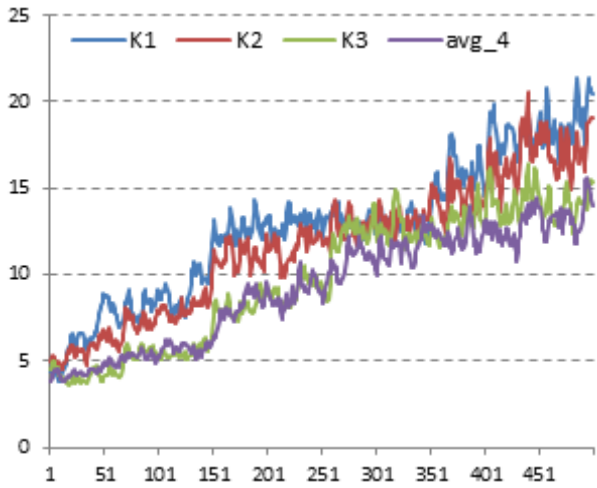

Fig. 10: Knowledge diffusion based on asymmetric settings

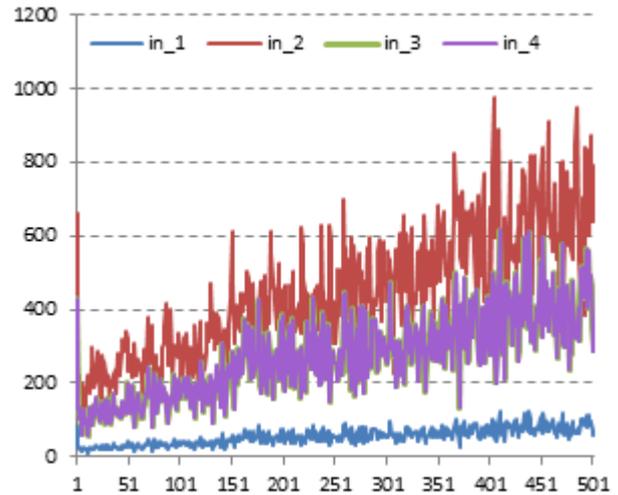

Fig. 11: Knowledge outgoing from vertical diffusion based on asymmetric relations

Parameters used for different relations between layers can be changed over time and it is related to the situation with emerging technologies and innovations. Obtained results showed diffusion processes between different layers. Dynamics of processes was simulated using both vertical and horizontal diffusion. Effect of deterioration was simulated as well as self-learning which results in changes over the time.

### 5.3.2 Modelling the role of experts

In the next step the role of experts within the network was modelled using asymmetric relations between knowledge components. Using the proposed model it is possible to simulate changes after adding experts with assigned knowledge higher than all network members. In the first stage of the simulation showed in Fig. 12 within 100 steps a stable result and equilibrium is visible. The proposed model can be used to estimate the effect of adding experts during the process. To improve the dynamics of diffusion expert can be added at a selected layer. After 100-th step of the simulation ten experts were added at first layer with maximal knowledge at the level of 50. It resulted in temporal growth within layer one and was followed by vertical diffusion to all layers. High increase of knowledge in a small segment of network resulted in deterioration process and reduction from average 18.45 down to 12.25 for layer number one.

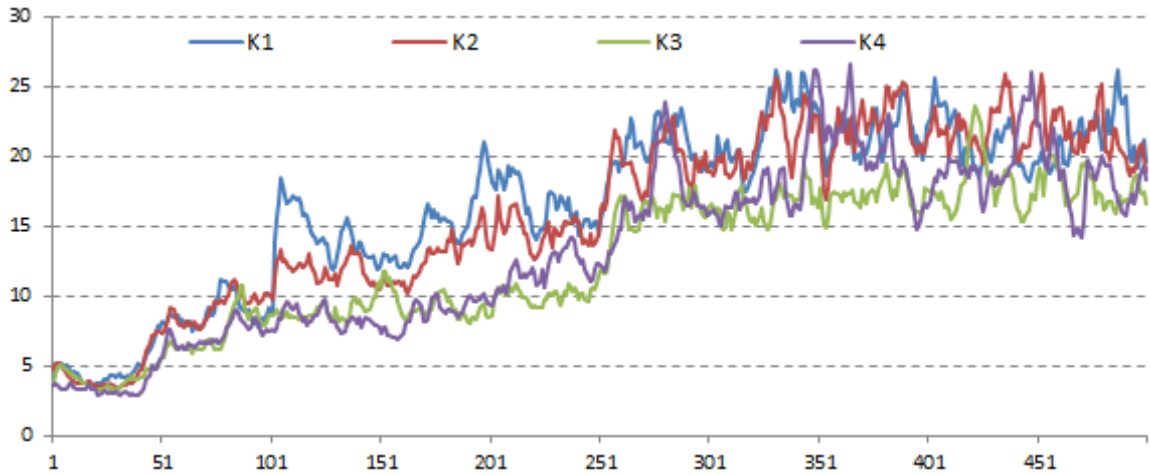

**Fig. 12: Multilayer knowledge diffusion improved by adding experts**

Knowledge transferred with vertical diffusion to other layer resulted a stable increase. In the next stage, 5 experts were added with maximal knowledge at layer one at the level of 25 and it was repeated after step 300. Adding continuously experts with smaller knowledge delivered better results than one time action based on experts with knowledge much higher than average knowledge within network. Activity for incoming and outgoing knowledge is showed in Fig. 13 and Fig. 14 respectively.

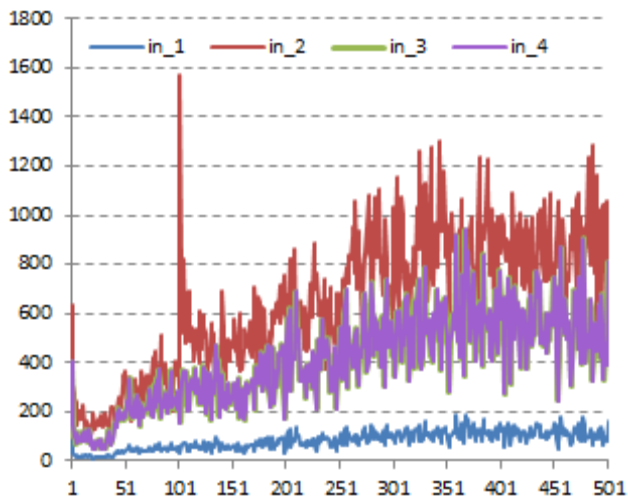

**Fig. 13: Knowledge incoming from vertical diffusion based on asymmetric relations**

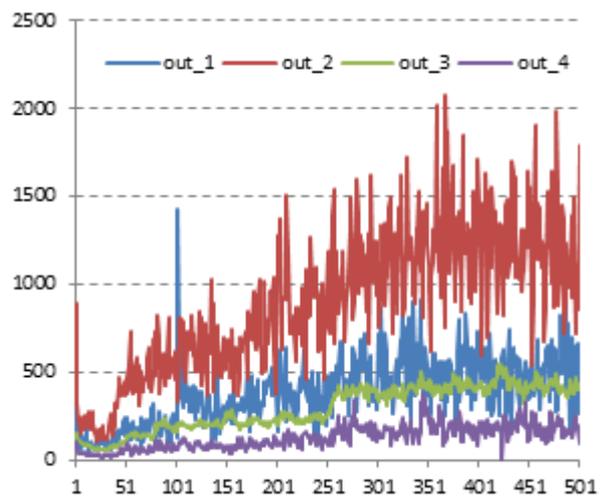

**Fig. 14: Knowledge outgoing from vertical diffusion based on asymmetric relations**

Using this approach it is possible to evaluate a better strategy to add a smaller number of experts with high knowledge or add higher number of experts with smaller knowledge. In the simulated conditions adding experts with high knowledge delivered worse results because observed deterioration process.

### 5.3.3 Modelling the changes in employment

Proposed model can be used for simulating situations of reduction of employment or job quitting. It was simulated in the next step and results are showed in Fig. 15. After 200-th step of the simulation 50 random employees were removed and 40% knowledge drop was observed. Improving this situation was possible After 300-th step where 10 experts with knowledge value at 50 were added at layer one and it helped to recover average knowledge.

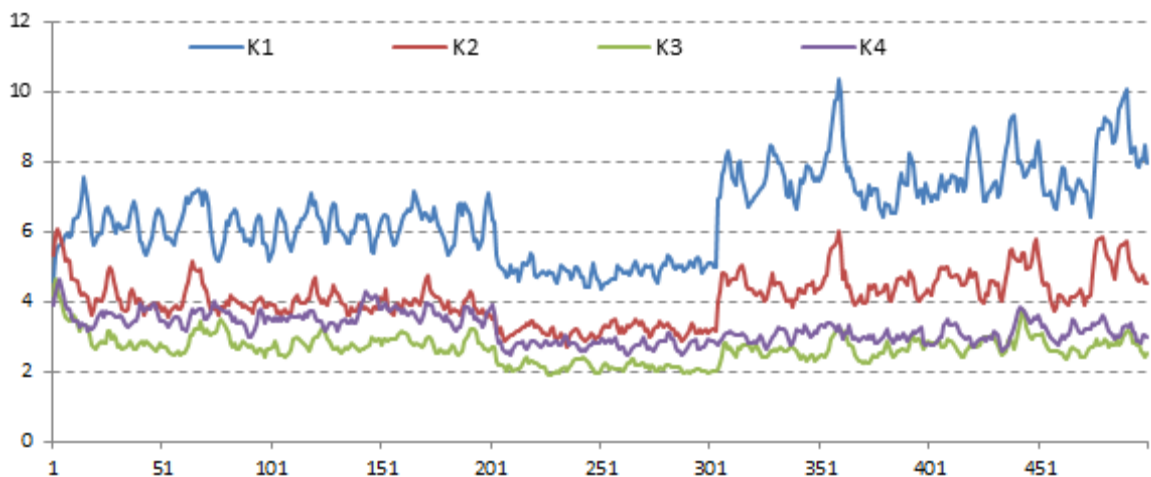

**Fig. 15: Multilayer knowledge diffusion with changes of employment**

Even though experts were added at single layer vertical diffusion helped to recover average knowledge at layer number two. Changes in employment are resulting different activity within incoming and outgoing knowledge at each layer what is illustrated in Fig. 16 and Fig. 17.

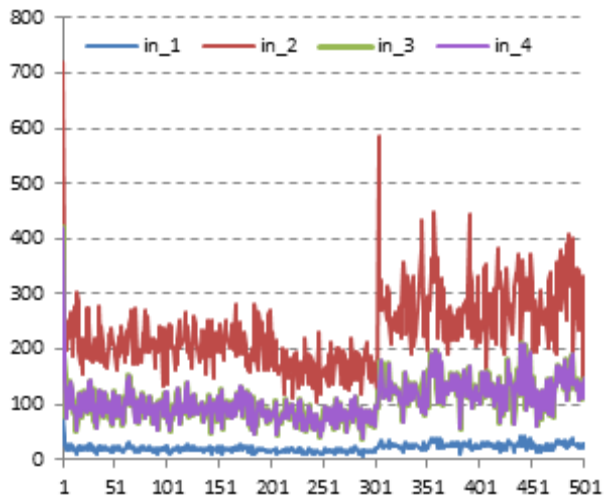

Fig. 16: Knowledge diffusion based on asymmetric settings

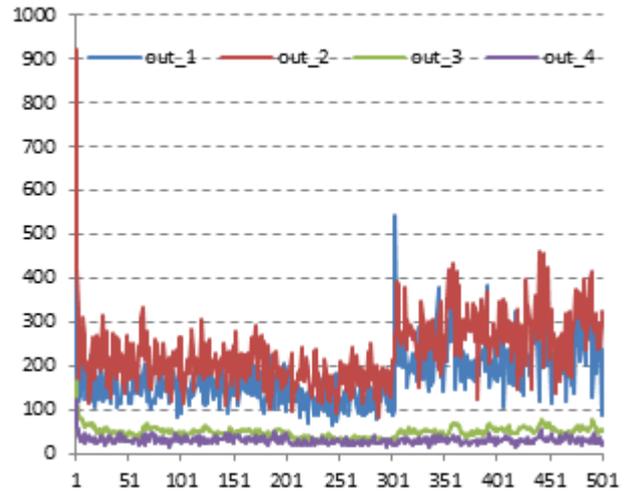

Fig. 17: Knowledge outgoing from vertical diffusion based on asymmetric relations

## 6  Discussion and conclusions

- Competence has a dynamic nature and can be represented by the set of its states variables values. The state variables represent different pieces of a person's competence thus the competence can be seen as a function of several time-based arguments, such as [22]: time of training (competence's acquiring), time of working actively using competence (competence's strengthening), time of inactivity (competence's decline), time of team work/problem solving (competence's transfer). In the future, such time-relevant arguments can create another dimension for diffusion in the model.
- The model has several properties that allow to manage the competence on an operational level. Current trend in literature is that competence management can be organized according to four kinds of mutual related processes [22]: competence identification, competence assessment, competence acquisition, competence knowledge usage. The proposed model supported all these processes on knowledge network's level:
    - competence identification: in order to create competence matrix all competence components have to be identified. As a result all competence's component are

recognized and measured. Moreover, based on the each layers analysis the trend related to the competence are observable and can be predicted.

- o competence assessment: due to all layers identification and description the assessment tools can be selected effectively.
- o competence acquisition: the diffusion mechanism supports the competence acquisition and help with the selection process of employee for training.
- o competence knowledge usage: the analysis of structure and content of network, behind the multilayer graph, give possibility to recognise the communities of practice.

- The model of knowledge diffusion process, which is based on vertical and horizontal diffusion and forgetting/self-learning processes gives a better picture of knowledge processing in an organization than the models from the literature. Proposed diffusion model allows to check what happens with competence value in an organization if, for example, the following scenario happens: base knowledge of all employees or only a few (experts) is increasing, some employees are removed from the organisation, some employees are transferred to other part of the organisation structure (new location in network). Moreover, we can used dedicated algorithms (e.g. [107]) to community detection in networks.

- Application of the theory of the Knowledge Spaces allows to estimate the level of knowledge in the context of existing and required competencies and relations between knowledge layers. We can precisely determine what part of worker's knowledge has to be increasing in order to achieve the required level of personal competence. The same problem for organisation required some optimisation approach. The optimisation problem is following: how to maximise the competence level on worker or organisation level regarding to social and cognitive personal worker's abilities, knowledge distribution and domain's structure whereas the constrains are time and cost of training?

- In the presented model, only two workers' roles are distinguished regarding knowledge level: expert and normal worker. However, it is possible to recognise more roles in order to model complex organisation structure. In the work [108] four main actor roles in knowledge workers community are recognised: the knowledge engineers (modelled the technical system part, wide expertise in the modelling and analysis of formal semantics), the core domain experts (excellent overview of the relevant topics and players in the

- domain), the domain experts (professionals representing specific domain), the application committee (works on the application level).
- In future approach to the modelling of the discussed issue it will be possible to change the relationship between the layers based on time-dependent function, or semantic relations reflecting business rules. In the presented approach, there are linear relationships between the layers described by vertical diffusion matrix.
- When the network of knowledge, competence and links is large the complexity of proposed approach is growing. The computational complexity depends on formulas for horizontal and vertical diffusion and self-learning/forgetting processes (2-10). If these formulas are nonlinear and mutually nested then the resources needed for calculations are significantly higher. The number of objects in the knowledge domain is not crucial due to formal nature of the Knowledge Space Theory (KST). In contrast to the semantic-rooted language (like OWL) in KST all relationships can be explicit interpreted and handled based on mathematical mechanisms. Moreover, the number of connection between workers (nodes) actually does not affect the whole approach, because the worker collaborated only with one other worker all the time.
- The notation of upper and lower shadow for worker's knowledge set gives opportunity to develop a cost estimation method for commence development. The cost estimation algorithm in the form of a group competences expansion algorithm is proposed in [109]. In this approach we have to recognise the acquired and required competence set and then based on the Competence Set Theory the cost of competence expansion is calculated.

# 7 Conflict of Interests

The authors declare that there is no conflict of interests regarding the publication of this paper.

# 8 Acknowledgments

The work was partially supported by Fellowship co-Financed by European Union within European Social Fund, by European Union's Seventh Framework Programme for research, technological development and demonstration under grant agreement no 316097 [ENGINE] and by The National Science Centre, the decision no. DEC-2013/09/B/ST6/02317.